\documentclass[
    ,final            
  ]
  {aipproc}
\usepackage[utf8]{inputenc}
\layoutstyle{6x9}

\begin{document}

\title{IceCube -- a new window on the Universe\footnote{
This paper is based on a talk presented by T. Gaisser on 1 September 2008 at the 3rd Latin American
School for Cosmic Rays, Arequipa, Peru in which co-author, Denis Robertson, was a student.}
}

\author{Thomas K. Gaisser }{
  address={Bartol Research Institute and Department of Physics and Astronomy\\
University of Delaware, Newark, DE 19716, USA}
}
\author{for the IceCube Collaboration}{
   address={\url{http://icecube.wisc.edu}}
}
\author{Denis Robertson}{
  address={Grupo de Física Fundamental, Facultad de Ciencias\\
Universidad Nacional de Ingeniería, Lima, Perú}
}

\begin{abstract}
 This paper gives an overview of the scientific goals of IceCube with an emphasis
on the importance of atmospheric neutrinos.  Status and schedule for
completing the detector are presented.
\end{abstract}

\maketitle


\section{Introduction--The IceCube Observatory}

\begin{figure}[htp]
  \includegraphics[width=10cm]{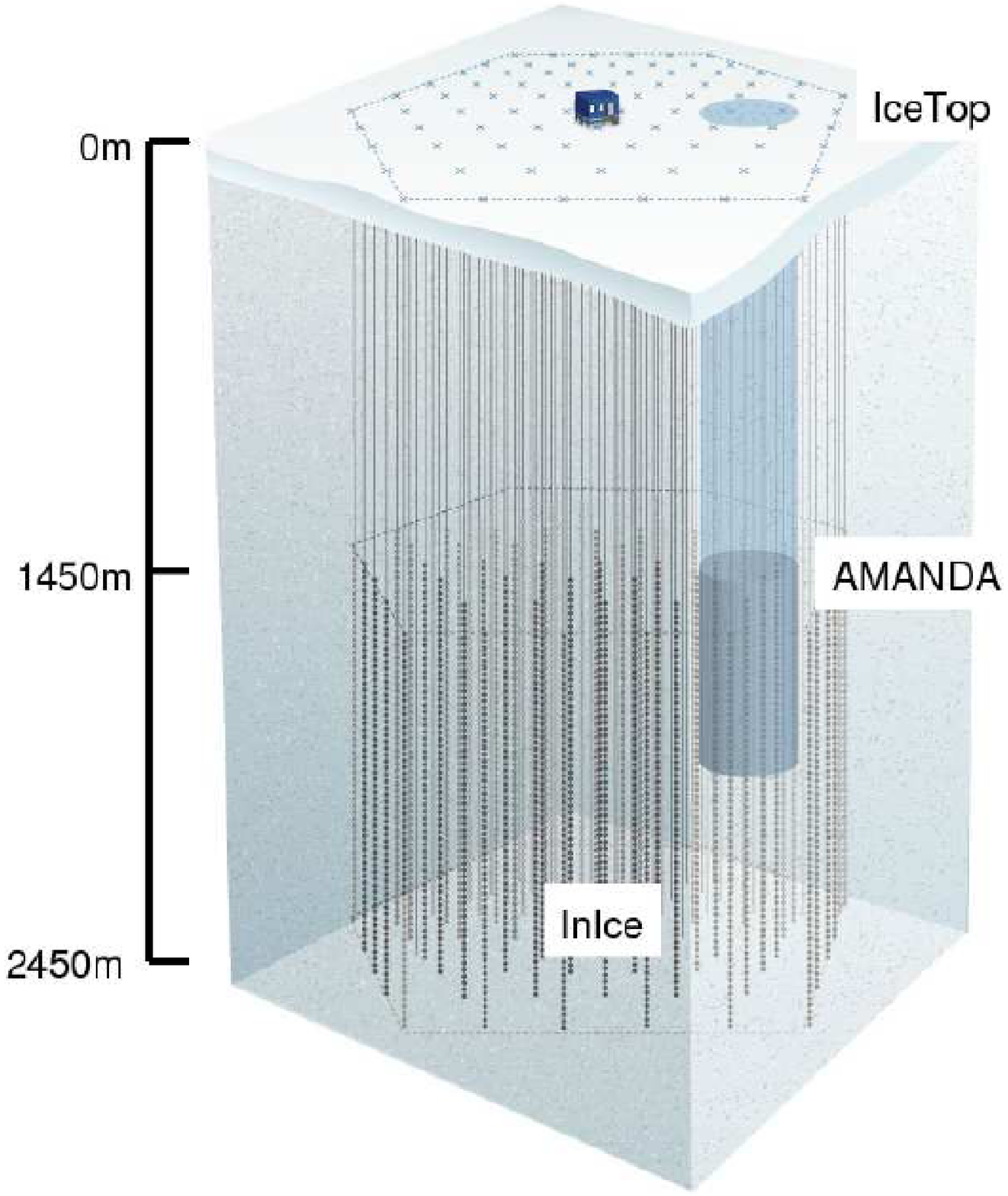}
  \includegraphics[width=6cm]{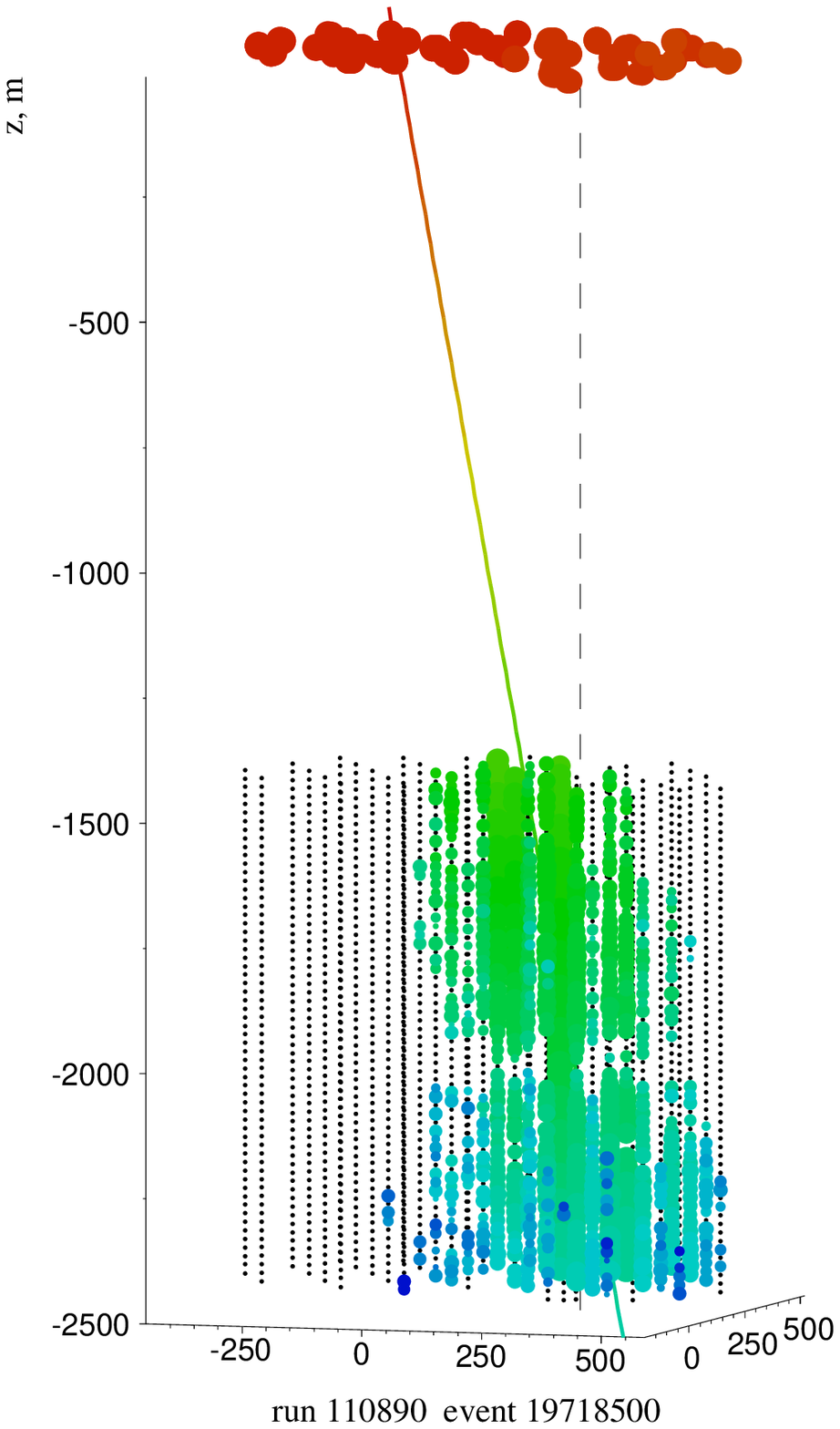}
  \caption{Artist's drawing of the IceCube detector.  IceCube currently has three 
components: the deep neutrino telescope, IceTop, and AMANDA, the smaller forerunner of 
the IceCube neutrino telescope.  On the right is a display of an EeV event recorded by
both IceTop and the deep array of IceCube in its present 40-string configuration.}
\label{BigEvent}
\end{figure}

The primary goal of IceCube is to detect high-energy neutrinos of extraterrestrial origin. Its method is to use the Earth as a filter by identifying upward moving events in the intense background of downward, cosmic-ray induced muons.  The
IceCube design (see Fig.~\ref{BigEvent})
calls for 4800 digital optical modules (DOMs) in the deep ice. 
There are 60 DOMs per cable separated from each other by 17 meters and deployed 
between 1450 and 2450 m into vertical holes melted in the ice. 
Each down-hole cable (called a "string") is separated by 
approximately 125 m from the neighboring strings. Each DOM is a glass sphere 33 cm in diameter capable of withstanding the high pressure of more than two km of water when deployed. Inside is a 25 cm Hamamatsu photomultiplier tube (PMT) and an electronics board that digitizes the signals locally using an on-board computer. There is also a local quartz clock that is synchronized to an accuracy of 3 ns to a single IceCube GPS clock on the surface~\cite{NIM}.

In addition to its deep neutrino telescope, IceCube also includes a surface air-shower array called IceTop. The surface array consists of stations located near the top of each in-ice string. Each station consists of two tanks separated from each other by 10 meters, and each tank is instrumented with two DOMs embedded in the clear ice that fills the tank.  IceTop DOMs are fully integrated into the IceCube data acquisition system. Together the surface array and the deep detectors of IceCube form a three-dimensional cosmic-ray detector.

The principle of particle detection is the same for in-ice and surface detectors; namely, to record the pulse of Cherenkov light from relativistic particles moving faster than the speed of light in ice, which has an index of refraction of 1.3.

When it is complete IceCube will instrument a cubic kilometer of ice with DOMs,
as illustrated in Fig.~\ref{BigEvent}a. The partial detector running with 40 strings
already extends a full kilometer in one direction. The kilometer scale of IceCube is set by the requirement that the detector is large enough to measure astrophysical neutrinos if they are produced with an energy density comparable to the energy density of ultra-high-energy cosmic rays~\cite{GHS}. Such a situation would naturally occur if accelerated particles are colliding with photons in the radiation fields of their sources in such a way that there is an equilibrium between production of secondary pions and neutrons. In such a case, the neutral pions decay to produce high-energy gamma rays, charged pions produce neutrinos and neutrons escape to contribute to the pool of ultra-high-energy cosmic rays.

Different neutrino flavors can be distinguished by characteristic patterns of light in the detector. Charged current interactions of muon neutrinos produce a muon which on average carries about 80\% of the neutrino energy. Such a muon has a long track so upward-moving muons that start outside the detector can also be detected if the neutrino is headed toward the detector.  The IceCube configuration is optimized for high-energy $\nu_\mu$-induced muons for this reason. The characteristic signal of an electron neutrino is a 5 - 10 meter long cascade produced by the high-energy electron from $\nu_e + N\rightarrow e + X$.  The signature of a $\nu_\tau$ is more complex and depends in detail on the energy of the neutrino and where it interacts. 

Observation of neutrinos from astrophysical sources will open a new window
because neutrinos have very small cross sections and can therefore emerge from deep
inside their sources and travel astronomical distances freely.
Of special interest are
regions of high density, such as microquasars in our galaxy, and
extra-galactic Active Galactic Nuclei (AGN) and Gamma-Ray Bursts (GRBs).
In such objects high gravitational forces associated with accretion of
magnetized plasma generate high energy beams of particles that can interact with the matter and photons in or near the sources and produce pions and other mesons.
Neutrinos will be produced from decays like
 $\pi^{\pm} \rightarrow \mu^{\pm} \nu_{\mu}$ followed by $\mu^{\pm} \rightarrow e^{\pm} \nu_{\mu} \nu_{e}$.  At production there is a relation of 2:1:0 of $\nu_{\mu}$:$\nu_{e}$:$\nu_{\tau}$. However, after large distances neutrino oscillations change this ratio to 1:1:1, which means that an astrophysical flux of $\nu_{\tau}$ is expected.  Because
$\tau$-neutrinos are rarely produced in the Earth's atmosphere, their
detection would be a strong indication of astrophysical origin.

The paper is organized as follows:  
We first describe in a general way  
the expected signals from the different neutrino flavors 
in the detector and how the energy and direction would be reconstructed from them.
Next, we discuss
atmospheric muons and neutrinos as background and calibration tool for IceCube.
We then discuss the implications for detection of astrophysical neutrinos.
Finally, we summarize the other scientific objectives of IceCube and 
conclude with a description of the status and plans for completion of the detector.

\section{Event reconstruction}

Charged particles traveling faster than the speed of light in ice
produce Cherenkov light which illuminates the optical modules of IceCube.
Cherenkov photons
 produce signals which are digitized and time stamped in the DOMs.  The times and waveforms provide the basic information from which events are reconstructed. Maximum-likelihood fitting techniques are used to reconstruct the direction and
energy of each event~\cite{reco1,reco2}.  In general, the total integrated
signal from each event is larger for events that deposit a large amount of
energy in the detector and smaller for less energetic events.  Corrections for
location of the light source relative to the DOMs is essential to obtain
an estimate of energy.  Corrections include not only the absolute distance
of the light source from each DOM but also the properties of absorption and
scattering in the ice as a function of depth.  In particular, there are several
dust layers, the most dense being at a depth of approximately 2100~m.

Flavors can be differentiated by their topologies:

\noindent
{\bf Electron Neutrino:} A $\nu_{e}$ interaction in the ice produces an electron which initiates an electromagnetic shower. The rest of the neutrino's energy goes into fragments of the target that produce a hadronic shower. The size of the showers is of order 10~m, which is small compared to the string spacing.  Each electron emits light at the $41^\circ$
Cherenkov angle, and there is a characteristic range of angles of the shower electrons
relative to the direction of the event.  Because of the small ratio of the region 
of the light source to the 
string spacing, it is difficult to reconstruct the direction of electron neutrinos.
On the other hand, energy reconstruction is relatively robust because the events
are contained inside the detector.
The Cherenkov light generated spreads over a volume which increases with energy (for example a sphere of radius 130 m at 100 TeV \cite{Hal}). 

\noindent
{\bf Muon Neutrino:} Charged current interactions of $\nu_{\mu}$ produce secondary muons which can travel long distances in ice (kilometers at TeV), generating showers along their track by bremsstrahlung, pair production and photo-nuclear interactions
if their energy is high enough. The energy of the muon decreases as it propagates in the ice, so the energy of the secondary showers diminishes.  Therefore the pattern of the light generated by a muon will be a Cherenkov cone along the track characteristic
of a minimum ionizing muon on which is superimposed bursts of light from
stochastic processes that on average decrease in intensity as the muon loses energy.
A minimum
ionizing muon can typically be seen by DOMs within 30 to 40 m of its track.

\noindent
{\bf Tau Neutrino:} $\tau$-neutrinos originate cascades very difficult to distinguish from electron neutrino cascades at low energies. However their flavor can be identified at higher energies. A possible favorable signature is the ``double bang'' in which two main cascades are produced, the first in the production of the tau lepton and the second in its decay. The tau also generates Cherenkov light during its life.  The energy window
for double bang events is constrained by the kilometer scale of
the detector to $5 < E_{\nu_{\tau}} < 20$~PeV so that the pathlength
of the $\tau$-lepton is between $0.25$ and $1$~km.  Other signatures could be the ``lollipop'' or the ``inverted lollipop''. In the first the tau lepton is produced outside IceCube so the hadronic cascade is not visible but the track of the tau and its decay are detected. The latter is the opposite case, the production of the tau occurs inside IceCube but not its decay. 
Ref.~\cite{Cow} gives a more detailed description of $\nu_{\tau}$ signatures 
in IceCube. 


\section{Atmospheric neutrinos}

\begin{figure}[htb]
\includegraphics[width=10cm]{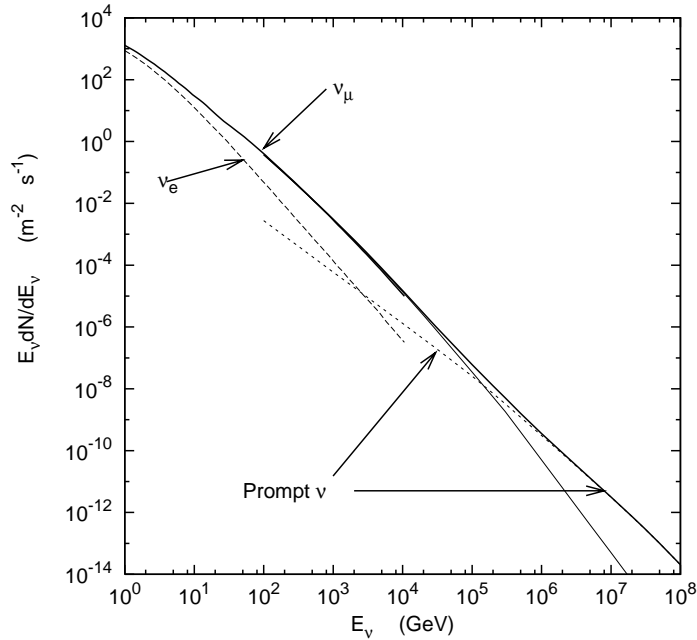}
\caption{Flux of atmospheric neutrinos integrated over all directions from below
the horizon.  The figure shows the sum of 
neutrinos and anti-neutrinos (adapted from
Figure~25 of Ref.~\cite{GH}).  The prompt component shown here corresponds to the
model of charm production in Ref.~\cite{Bugaev} (see text).}
\label{atmosnu}
\end{figure}

The main event rate in IceCube 
is from atmospheric muons created by cosmic-ray interactions in the atmosphere. The background events are about 1 million times more numerous than events generated by atmospheric neutrinos.  Understanding the downward flux of atmospheric muons
and reconstructing the muon directions is a prerequisite for analysis of IceCube
data. 

To eliminate the very high background of downward muons, IceCube uses upward-going muons to identify and
study neutrino interactions.  The Earth is used as a filter to select 
muons produced by charged current interactions of muon neutrinos in the ice or rock below or beside IceCube. After upward moving events are selected, there are still more background events than signal. After the first cut(s) the background consists of misreconstructed downward events and events in which two unrelated muons go through the detector within the same event time window.  Further cuts are made to reduce these backgrounds and arrive at the level where most of the upward events are
from atmospheric neutrinos.

The intensity of atmospheric neutrinos is fairly well known into the TeV region.
It decreases with energy at a higher rate than is expected for astrophysical neutrinos, falling approximately like E$^{-3}$ and steepening to E$^{-3.7}$ for E $\gg$ 1 TeV. 
Figure~\ref{atmosnu} showing the total atmospheric neutrino spectrum is 
adapted from figure 25 of 
Ref.~\cite{GH}. For energies higher than 100 GeV most of the neutrinos come from decay of kaons until the energy is so high that neutrinos from charm dominate~\cite{Tokyo08}. 
At high energy the flux is higher at larger zenith angles and up-down symmetric 
except for the
few GeV range where geomagnetic effects play a role.  
The uncertainties in the fluxes are around 15\% at $1$~TeV,
and they increase with energy.  Uncertainties in flavor ratios and in
angular dependence of the neutrino spectra are significantly less than
uncertainties in the normalization~\cite{BRGS}. 

The level of charm production becomes a significant
uncertainty at high energy.  The model for charm~\cite{Bugaev}
shown in Fig.~\ref{atmosnu} is near the maximum allowed~\cite{Gelmini} by current
observations.  A new calculation of charm production~\cite{ERS} is an order of magnitude lower.  Because charm contributes an isotropic ``prompt" component
of neutrinos with a harder
spectrum than neutrinos from kaons and pions, it is potentially a 
significant background in the search for a diffuse flux of astrophysical neutrinos
above 100 TeV.

\subsection{Response of IceCube to atmospheric neutrinos}

During construction, IceCube is augmented with new strings each Austral
summer (November - February).  A new science run begins in April after the new DOMs
are frozen in and continues until the following March.  From April 2007 to
March 2008 IceCube ran with 22 strings in the deep ice and with 26 stations of
IceTop on the surface.  
IceCube has been running in a 40 string configuration since April, 2008.
Analysis of data from IceCube-22 (IC22) is currently well advanced.
The total exposure of IC22 (km$^2$-sr-years) 
is already comparable to seven years of AMANDA data, and
a discussion serves to illustrate the methods and first results of IceCube.
\vfill\eject

\vspace*{-.5truecm}
\begin{figure}[thb]
  \includegraphics[width=8.5cm]{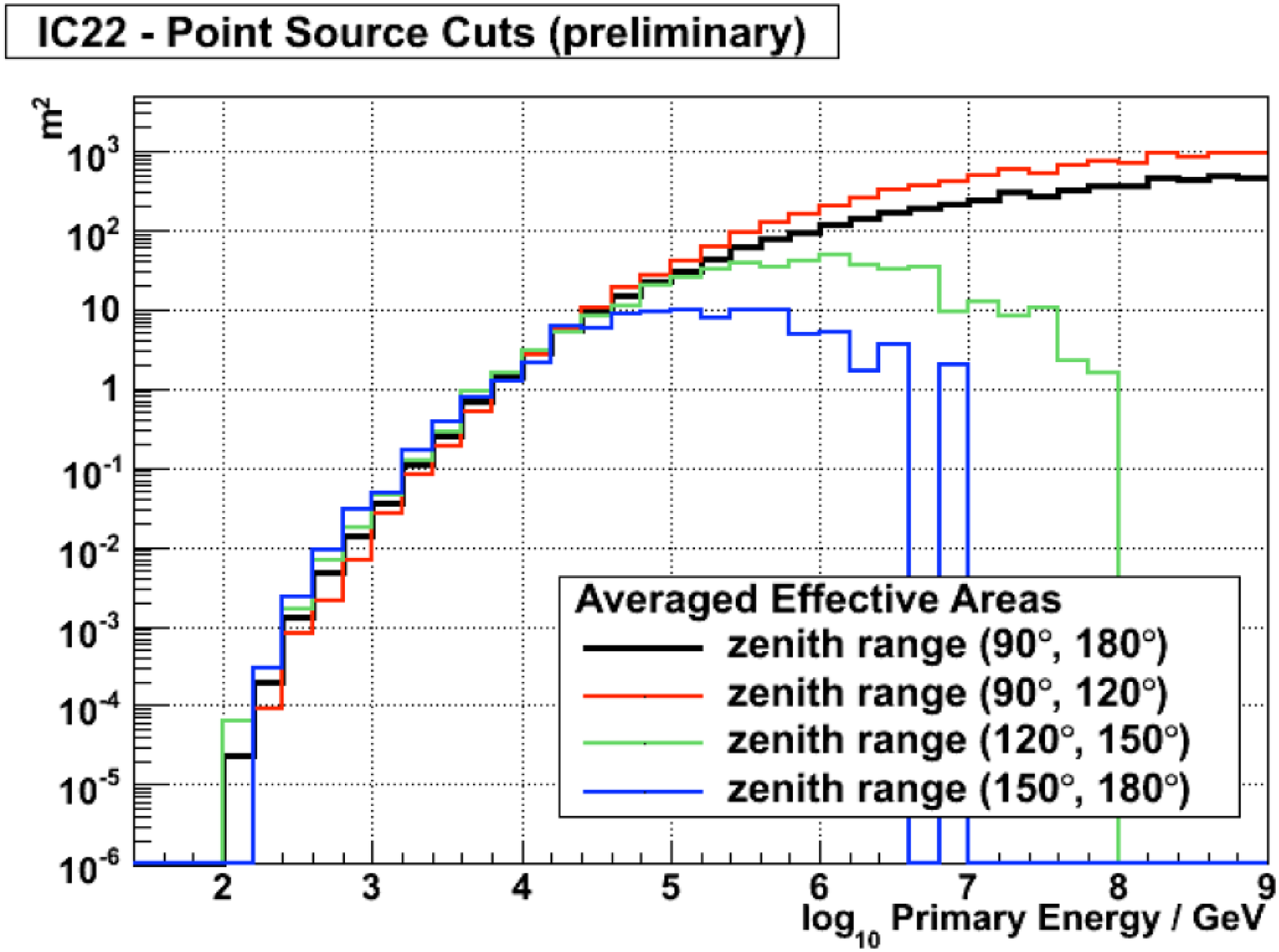}
  \caption{Neutrino effective area of IceCube22~\cite{Karle} averaged over all directions from below (heavy line).  Light lines average over angular bins from horizontal (highest) to vertical (lowest).}
\label{Aeff22}
\vspace*{-.5truecm}
\end{figure}

The neutrino effective area is defined so that the product of neutrino intensity
multiplied by the effective area gives the event rate.  Its value as a function
of angle and energy for IC22 is shown in Fig.~\ref{Aeff22} from Ref.~\cite{Karle}.
The neutrino effective area is given by
\begin{equation}
A_{\rm eff}(\theta,E_\nu)\;=\;\epsilon(\theta)\,A(\theta)\,P(E_\nu,E_{\mu,{\rm min}})
\,e^{-\sigma_\nu(E_\nu)N_A\,X(\theta)},
\label{Aeff}
\end{equation}
where $\epsilon(\theta)$ is the efficiency for a detector of projected area
$A(\theta)$ to detect a muon incident at zenith angle $\theta$.  The exponential
expresses the muon attenuation in the Earth for angle $\theta$ below
the horizon, where $X(\theta)$ is the amount of matter (g/cm$^2$) along
the chord through the Earth.  The factor $P(E_\nu,E_{\mu,{\rm min}})$ is the
probability that a muon neutrino on a trajectory that will intercept the
detector gives a visible muon in the detector.  It is
\begin{equation}
P(E_\nu,E_{\mu,{\rm min}})\,\,=\,\,N_A\,\int_{E_\mu,{\rm min}}^{E\nu}\,{\rm d}E_\mu
{{\rm d}\sigma_\nu\over {\rm d}E_\mu}\,R(E_\mu,E_{\mu,{\rm min}})\, ,
\label{nuprob}
\end{equation}
where the integrand is the product of the charged current differential cross section
and $R$ is the average distance traveled by a muon with energy $E_\mu$ at production
before its energy is reduced to $E_{\mu,{\rm min}}$, the minimum energy required
upon entering the detector for the event to be reconstructed well.  The additional 
contribution from neutrinos that interact within a large detector can be computed
in a straightforward way and added as an extra contribution.

\begin{figure}[htb]
\includegraphics[width=8cm]{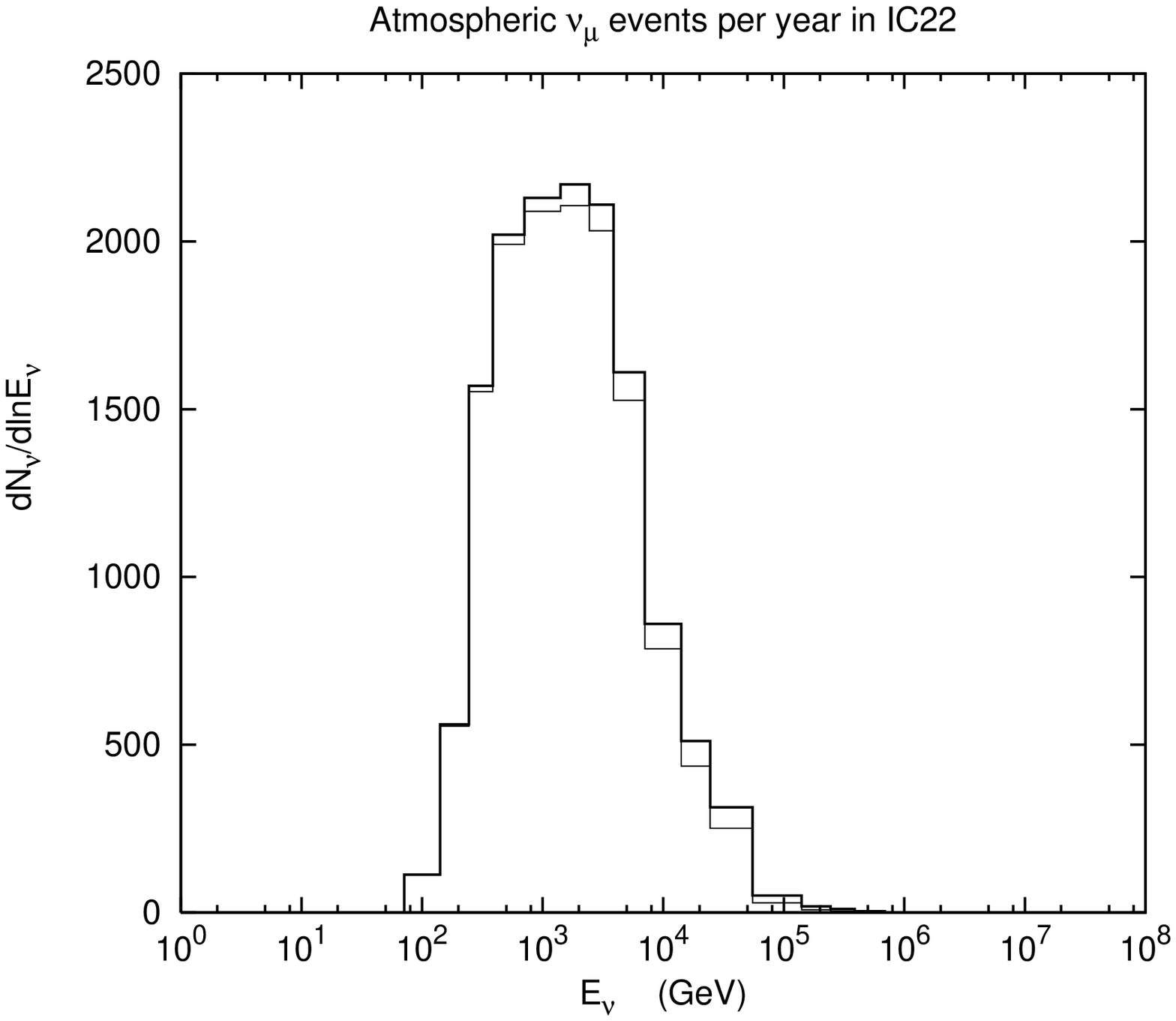}
\includegraphics[width=8cm]{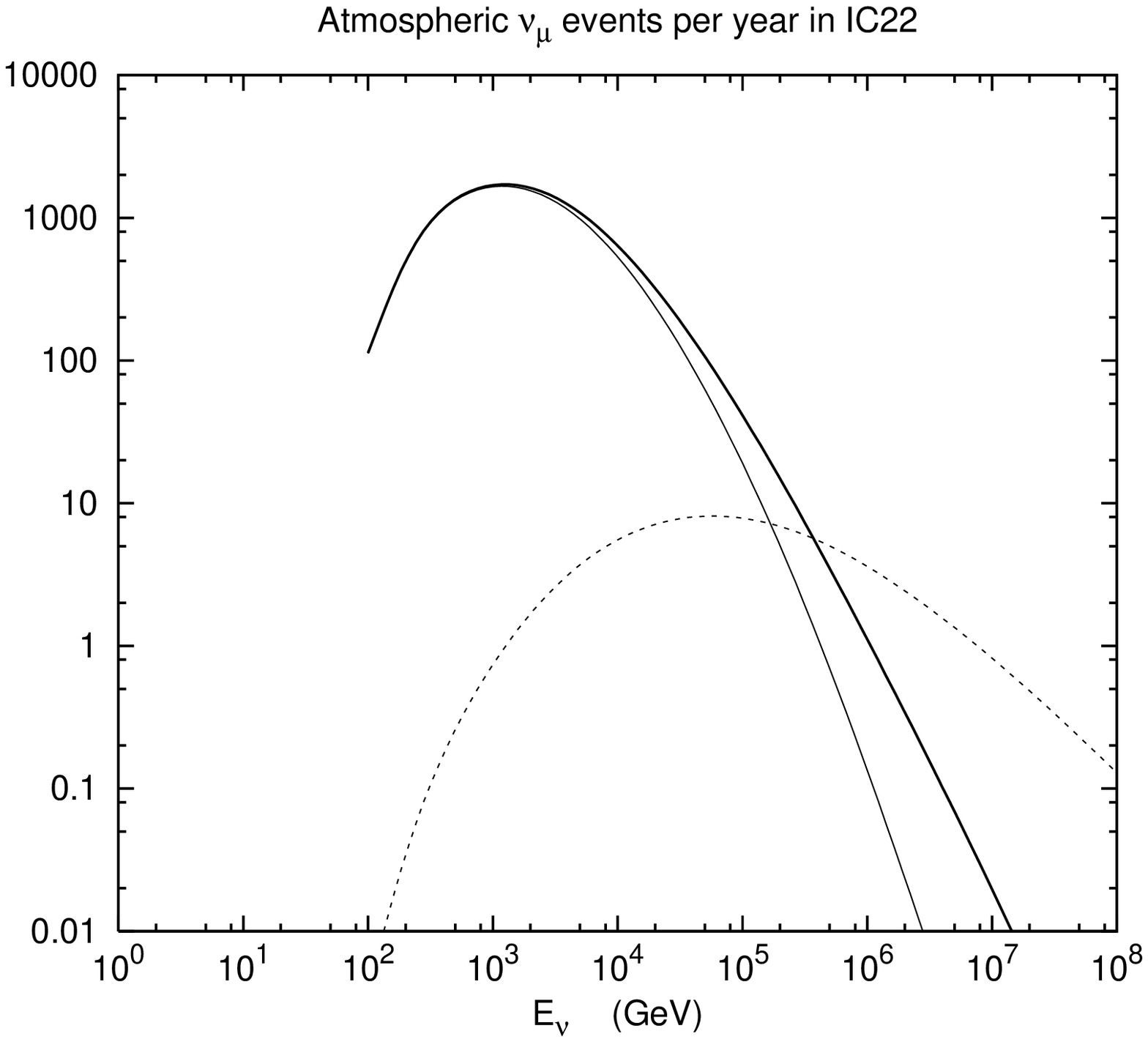}
\caption{Response of IC22 to atmospheric muon neutrinos (left: linear scale;
right: logarithmic scale).  The upper line includes prompt neutrinos from decay
of charmed hadrons normalized for the model of Ref.~\cite{Bugaev}.  
The broken line shows the response for
an $E_\nu^{-2}$ differential neutrino spectrum (see text).  There are about 6000 
well-reconstructed atmospheric neutrinos per year in IC22.
}
\label{response}
\end{figure}

The product of the effective area for the lower hemisphere (heavy histogram 
in Fig.~\ref{Aeff22}) and the atmospheric neutrino flux from below the
horizon (Fig.~\ref{atmosnu})
gives the differential response of IC22 to atmospheric $\nu_\mu$.
This response is shown in Fig.~\ref{response}.
The dashed line in the logarithmic plot (Fig.~\ref{response}b) 
also shows the response of IC22 to a potential
diffuse astrophysical flux of muon neutrinos~\cite{WB} with a spectrum
\begin{equation}
E_\nu^2{{\rm d}N_\nu\over{\rm d}(E_\nu)}\,=\,2\times 10^{-4}\,{\rm GeV}
{\rm m}^{-2}{\rm s}^{-1}{\rm sr}^{-1}.
\label{WaxBah}
\end{equation}

\subsection{Muons in IceCube}

IceCube now has the largest instrumented volume for detecting
and reconstructing muons with energies in the TeV range and above.
The measurement of muons from all directions is therefore a major
benchmark for IceCube.  Because of the vast difference in rates
of downward atmospheric muons and upward neutrino-induced muons,
the measurement requires a dynamic range of sensitivity of more than
six orders of magnitude. 

\begin{figure}[hbt]
  \includegraphics[width=4in]{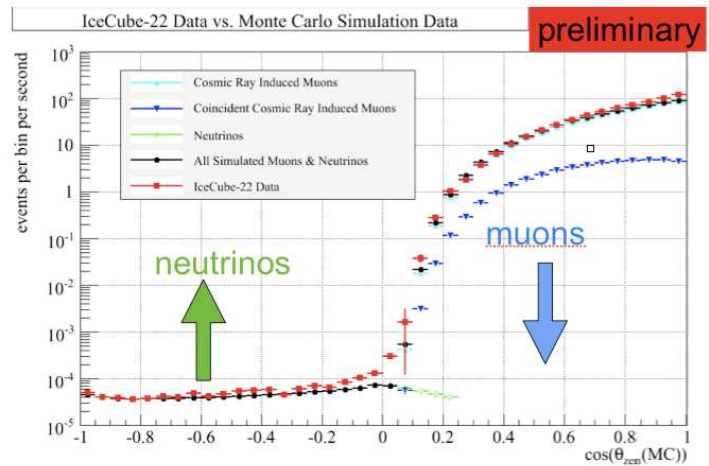}
  \caption{Reconstructed direction of muons in IceCube as a function of zenith angle
at trigger level~\cite{Berghaus}.
The upward hemisphere is populated mainly by neutrino-induced muons while the downward 
hemisphere is from muons produced in the atmosphere above the detector.  At trigger level
the cuts are looser than in Fig.~\ref{Aeff22}.}
\label{allsky}
\end{figure}

Figure.~\ref{allsky} shows the measured distribution of muons
in IC22 as a function
of zenith angle for the whole sky--atmospheric muons from above and
atmospheric neutrinos from below.  The data are compared with simulations
that show the components separately (including the contribution from
accidentally coincident downward muons).

\section{Scientific objectives of IceCube}

\subsection{Astrophysical neutrinos}

To study astrophysical neutrinos, we first need to discriminate in IceCube which events are caused by atmospheric and which ones by astrophysical neutrinos. One way to differentiate astrophysical neutrinos from atmospheric ones is by means of their energy spectrum,
as illustrated in Fig.~\ref{response}b.   This is a challenging quest because the 
expected intensity is low.
Another way would be to see several neutrinos from the same direction, especially if the direction is associated with a known source of high-energy gamma-rays. Even better would be to see two or more neutrinos at the same time and from the same direction as a gamma-ray burst.

\begin{figure}[hbt]
  \includegraphics[width=10cm]{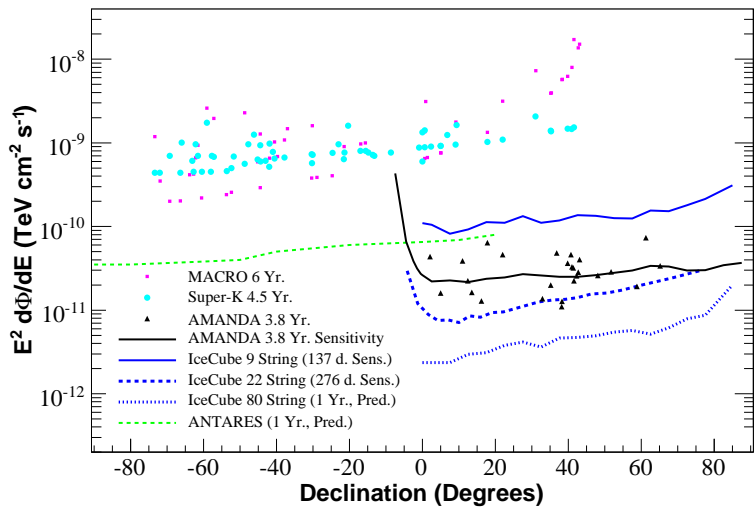}
  \caption{Limits on neutrinos from various potential sources~\cite{AMANDAlimits}.
Dots represent limits for individual sources.  Lines indicate predicted levels of
sensitivity for IceCube.}
\label{limits}
\end{figure}

Potential sources include
AGN and GRBs. Both of these are likely
to be cosmic particle accelerators.  If so, neutrinos could be produced when accelerated protons (or nuclei) interact with photons or gas near the sources. AGN and GRB are extra-galactic sources. Galactic cosmic-ray accelerators such as supernova remnants and
interacting X-ray binaries, or micro-quasars, might also produce neutrinos.  Limits 
from a search with AMANDA for Northern hemisphere point sources of neutrinos in
data collected over seven years (3.8 years of live time)
are shown in Fig.~\ref{limits}~\cite{AMANDAlimits}.  The blue lines show the
sensitivity of IceCube at various stages of construction.  IC22 is already
more sensitive than AMANDA, and one year of the full IceCube is more sensitive
by an order of magnitude.  Initial results of searches for point sources
with IC22 have been reported at conferences~\cite{BazoAlba,Berghaus}.

\subsection{Other objectives}

\subsubsection{Indirect detection of dark matter}
IceCube also looks for neutrinos as a signature of dark matter by looking for neutrinos from the Sun or the center of the Earth where weakly interacting massive particles (WIMPs) could accumulate and annihilate each other.  The indirect search for WIMPs in 
the Sun is complementary to direct searches because the capture rate of WIMPs
depends on their spin-dependent cross section for interaction with
protons in the Sun.  In contrast, direct searches with nuclei such as
xenon or silicon depend on coherent, spin-independent cross sections.
So the two approaches are sensitive to different regions of supersymmetric
parameter space~\cite{Hal2}.  Limits with AMANDA have been published~\cite{AMANDAWIMP},
and a search with IC22 is forthcoming.

\subsubsection{Search for exotic particles and processes}

An important signature 
to look for is any signal that corresponds to propagation
of a track through IceCube at less than the speed of light.  Possibilities
to generate such signals 
include massive magnetic monopoles, Q-balls and massive nuclearites~\cite{exotic}.

Another search looks for non-standard oscillations using the atmospheric 
neutrino beam.  As mentioned above, measurement of atmospheric neutrinos 
and reconstruction of their energy spectrum is a benchmark measurement for
IceCube.  In the process of making this measurement it is natural to
look for new physics that might show up in the neutrino spectrum at an energy
higher than what has previously been explored.  In theories that allow
violation of the principles of relativity theory, neutrino eigenstates
can exist with mixing effects that show up as energy increases (opposite
to the case for standard neutrino oscillations).  Analysis of the full 7-year
AMANDA data sample from this point of view provides new limits on such models
while extending the measurement of the atmospheric $\nu_\mu$ spectrum
into the multi-TeV energy range~\cite{Kelley}.

\subsubsection{Geophysics}
Given sufficient statistics that will be accumulated by IceCube on a
ten-year time scale, it may be possible to use the known angular dependence
of the atmospheric neutrino beam to probe the core-mantle density transition
deep in the Earth~\cite{Hal4}.  The Earth begins to absorb neutrinos from
directly below above $10$~TeV (see the $150^\circ$-$180^\circ$ line in
Fig.~\ref{Aeff22}).  Because of the steep decline of the atmospheric
neutrino signal above $10$~TeV (Fig.~\ref{response}a), good energy
resolution is required to exploit this possibility~\cite{Tokyo08}.

\subsubsection{Supernovae and Solar physics}
Monitoring the event rates in individual 
IceCube DOMs is the basis for two quite different
physics goals: supernova watch and solar and heliospheric physics.
By design IceCube is a coarse detector with a threshold
of order 100 GeV for muons to have a sufficiently long
track to be reconstructed.  Therefore, individual events of low energy
cannot be reconstructed.  However, when a sufficiently nearby 
supernova occurs the random interactions of anti-neutrinos
on protons producing $\approx 20$~MeV positrons near individual DOMs will
cause the overall counting rate of the detector to go up~\cite{Hal3}.
Typical counting rates of individual DOMs in the deep, dark
ice are in the vicinity of 250Hz after removing correlated afterpulses.
This low rate allows detection of a supernova
with good probability out to the Small Magellanic Cloud (62 kpc).

Counting rates in IceTop DOMs on the surface are higher, of order 
2 kHz, even though their discriminator thresholds are set higher than in-ice DOMs.
The steady, uncorrelated counting rate of the DOMs in IceTop tanks is
due to low-energy secondary cosmic ray electrons, photons and muons hitting
the tanks.  These particles are produced by the continuous flux of galactic
cosmic rays with energies of order 10 to 100 GeV interacting in the atmosphere.
When an energetic solar flare accelerates particles of several GeV
that reach the atmosphere and interact, the event shows up as an abrupt increase in
the counting rate of DOMs on the surface followed
by a gradual decline as the intensity of the flare particles decreases.  

The first extra-terrestrial
event seen with IceCube was the solar flare event of December 13, 2006 when there were
sixteen IceTop tanks in operation.  Using the fact that the increase in counting
rate in individual DOMs depends on its discriminator threshold, it was 
possible to extract some information about the spectrum of the particles
in the event~\cite{solar}.  The sensitivity provided by the full IceTop with
160 tanks is expected to provide an additional tool for solar and heliospheric
physics.

\subsubsection{Cosmic-ray physics with IceCube}

The main goal of IceCube as a cosmic-ray detector is to measure the energy spectrum and relative composition of protons and heavy nuclei of the primary cosmic rays from 1 PeV to 1 EeV.  The goal is to look for a signature of a transition from galactic cosmic rays to a population of particles from extra-galactic sources.  Preliminary results from
data taken in 2007 with 26 IceTop stations are promising.  Using the angular
dependence of air showers as a function of shower size on the ground, clear
evidence of sensitivity to primary composition is seen~\cite{Klepser}.  The 
sensitivity arises from the fact that showers generated by protons are more
penetrating than showers generated by heavy nuclei, so the proton component
contributes relatively more to the event rate at larger zenith angles.
This result is from IceTop alone.

Events with trajectories that pass through IceTop and the deep array
of IceCube offer another handle on composition.  The signal in IceTop
is primarily due to the electromagnetic component of the shower.  
This component is absorbed by the ice leaving only the penetrating
high-energy muons in the core of the shower that reach $>1.5$~kilometer
to produce signals in the deep IceCube DOMs.  
The ratio of muon component to electromagnetic
component depends on composition in a way that is complementary
to the measurement of angular dependence on the surface.  Requiring a consistent
interpretation of the two measurements will be a strong constraint on
the analysis, which depends on comparison to Monte Carlo simulations of
shower development.  

Other shower properties will also provide additional
constraints on the interpretation.  An example is the relative content
of low-energy muons at large distances from the shower core on the
surface~\cite{Kolanoski}.

The acceptance of full IceCube for coincident events that trigger
both IceTop and the deep array
is $0.3$~km$^2$sr, which is large enough to see a few events
in the EeV range.  Figure~\ref{BigEvent}b shows one such event observed in 2008 
with IC40 when IceCube was half its design size.  Such an event 
contains some 2000 muons with sufficient energy to reach the deep detector.
Two important properties of the ice clearly show up in this event display.
One is the main dust layer in the middle of the deep strings (around 2100
meters below the surface).  The other is the exceptional clarity of the ice
below the dust layer, which is apparent from the large amount of light
and its extent.  This is despite the fact that some of the muons are
ranging out inside the detector so the source of light is decreasing with
depth in the detector.

\section{Status and plans}

IceCube is currently operating in its 40-string configuration with 40 IceTop
stations.  This IC40 run began in April 2008 and will continue through
the 2008-2009 construction season until the end of March 2009.
The current 2008-2009 deployment season saw the addition of 19 
IceTop stations and 19 more strings.  A new run will begin with the
newly deployed strings and tanks in April 2009 after the new DOMs have
been commissioned.

One of the new strings installed in the current season is
a specially configured ``deep core" cable with 50 closely spaced,
high quantum efficiency (HQE) DOMs below the dust layer
and 10 DOMs above the main dust layer.  The concept is to build
a densely instrumented, deep subarray within IceCube that will
replace AMANDA~\cite{Resconi}.  Six densely instrumented strings will be placed
around a central string of the original plan, equidistant
between the central cable and nearest six surrounding cables of the original
plan.  These 13 strings (six specially configured and seven with standard
DOM spacing) will constitute the deep core subarray of IceCube.  Its
location is such that there are at least three rings of IceCube strings
on the standard 125 meter grid between the deep core and the edge of the
array.  This deep, inner core will be surrounded laterally and above by
some 4500 standard IceCube DOMs.  The plan is to turn off AMANDA at the
end of the current run (end of March 2009) and to install the remaining
five special strings of the inner core in the 2009-2010 deployment season.

The deep inner core subarray will increase the sensitivity of IceCube
at low energy ($<\,100$~GeV) allowing study of neutrino oscillations~\cite{Rott}
and increasing the sensitivity for indirect searches for WIMP annihilations
in the Sun.  
Using the outer veto area will enable IceCube to move toward the
regime of much more densely instrumented detectors like Super-K.
The goal is to use the veto capability to identify a class of 
partially ``contained" events
in which the vertex is known with high probability to be
inside the deep core fiducial volume.  This would allow the identification of
a fraction of the neutrinos from above as well as those from below.

The possibility of increasing the reach of IceCube at higher energy
is also under consideration.  A straightforward way to accomplish this
would be to place some of the last strings of IceCube further out from
the center~\cite{Karle}.  The goal of such an extension would
be to increase coverage in the PeV neutrino energy range 
so the detector would have better sensitivity to weak astrophysical
signals with hard spectra above the atmospheric neutrino background. 
The goal of measuring the intensity of cosmogenic
neutrinos (neutrinos produced by interactions of ultra-high-energy
cosmic rays with the cosmic microwave background radiation) may be beyond
the reach of IceCube itself.  New techniques for covering much larger
target volumes are being explored with test devices deployed in
IceCube holes.  These include radio and acoustic techniques as
discussed in Ref.~\cite{ARENA}.

The final deployment season for IceCube is planned for 2010-2011.
The plan is to manage resources well enough to be able to deploy
the 80 standard IceCube strings of the original plan as well as
the 6 special strings of the deep core.


\begin{theacknowledgments}
We are grateful to the conference organizers for the opportunity to 
participate in the 3rd Latin American School for Cosmic Rays.
The IceCube Collaboration acknowledges support from the following agencies:
 U.S. National Science Foundation-Office of Polar Programs, U.S. National Science Foundation-Physics Division, 
U. of Wisconsin Alumni Research Foundation,
 U.S. Department of Energy, NERSC, the LONI grid; Swedish Research Council, K. \& A. Wallenberg Foundation, Sweden; German Ministry for Education and Research, Deutsche Forschungsgemeinschaft; Fund for Scientific Research, IWT-Flanders, BELSPO, Belgium; the  Netherlands Organisation for Scientific Research.
\end{theacknowledgments}

\end{document}